\documentclass[
               ,final 
              ]{aipproc}
\layoutstyle{6x9}

\usepackage{natbib}

\begin{document}


\title{The Burst Mode of Accretion in Primordial Star Formation}

\author{Alexander L.~DeSouza}
       {address={Department of Physics \& Astronomy, Western University, London, Ontario, N6A 3K7, Canada}}
\author{Eduard I.~Vorobyov}
       {address={Institute for Astrophysics, The University of Vienna, Vienna, 1180, Austria},altaddress={Institute of Physics, Southern Federal University, Stachiki 194, Rostov-on-Don, 344090, Russia}}
\author{Shantanu Basu}
       {address={Department of Physics \& Astronomy, Western University, London, Ontario, N6A 3K7, Canada}}

\begin{abstract}
We present simulation results for the formation and long-term evolution of a primordial protostellar disk harbored by a first star. Using a 2+1D nonaxisymmetric thin disk numerical simulation, together with a barotropic relation for the gas, we are able to probe ${\sim}20~\mbox{kyr}$ of the disk's evolution. During this time period we observe fragmentation leading to loosely bound gaseous clumps within the disk. These are then torqued inward and accreted onto the growing protostar, giving rise to a burst phenomenon. The luminous feedback produced by this mechanism may have important consequences for the subsequent growth of the protostar.
\end{abstract}



\maketitle


\section{Introduction}

Although the existence of protostellar disks is a ubiquitous outcome of the present-day star formation process, the importance of these structures in the evolution of the first stars in the early universe has come to light only recently \citep{stacy2010,clark2011}; with their significance not fully understood. Numerical simulations by several authors have converged upon the idea that collapsing primordial cores produce a much more structure-rich environment than previously thought: binary pairs \citep{machida2008}, embedded cluster formation \citep[e.g.,][]{clark2008}, and vigorous small-scale fragmentation \citep{greif2012}, are all likely outcomes of primordial disk formation and fragmentation.

A major limitation of the aforementioned calculations however, is the inability to follow the disk evolution for much more than a thousand years. Herein we present 2+1D numerical simulations of the formation and evolution of a primordial circumstellar disk over time scales that are physically relevant to the global picture of Population III protostellar evolution.


\section{Numerical Model}

We employ a version of our 2+1D numerical hydrodynamic simulations \citep[see][for details]{vorobyov2010}, with appropriate modifications introduced for star and disk formation in the early universe. In particular, the hydrodynamic equations are closed with a barotropic relation that we derive from the detailed 1D chemical and thermal calculations of \citet{omukai2005}. A radially and azimuthally varying vertical scale height is determined using the assumption of local hydrostatic equilibrium.

The simulation contains approximately $200~\mbox{M}_{\odot}$ of material, initialized with a uniform temperature of $250~\mbox{K}$, within a simulation region of radius $0.4~\mbox{pc}$. This is representative of a typical Jeans-mass excess of gas collapsing within a virialized dark matter mini-halo \citep{yoshida2006}. The initial radial profiles of the surface mass density and angular velocity consist of a flat central region that then asymptotes to an $r^{-1}$ profile. The simulation is terminated once approximately $30~\mbox{M}_{\odot}$ of material has been accreted onto the protostar. Our calculations do not include the effects of ionization, which become important as the star grows beyond this mass.


\section{Results \& Discussion}

\begin{figure}[t!]
  \includegraphics[width=0.333\textwidth]{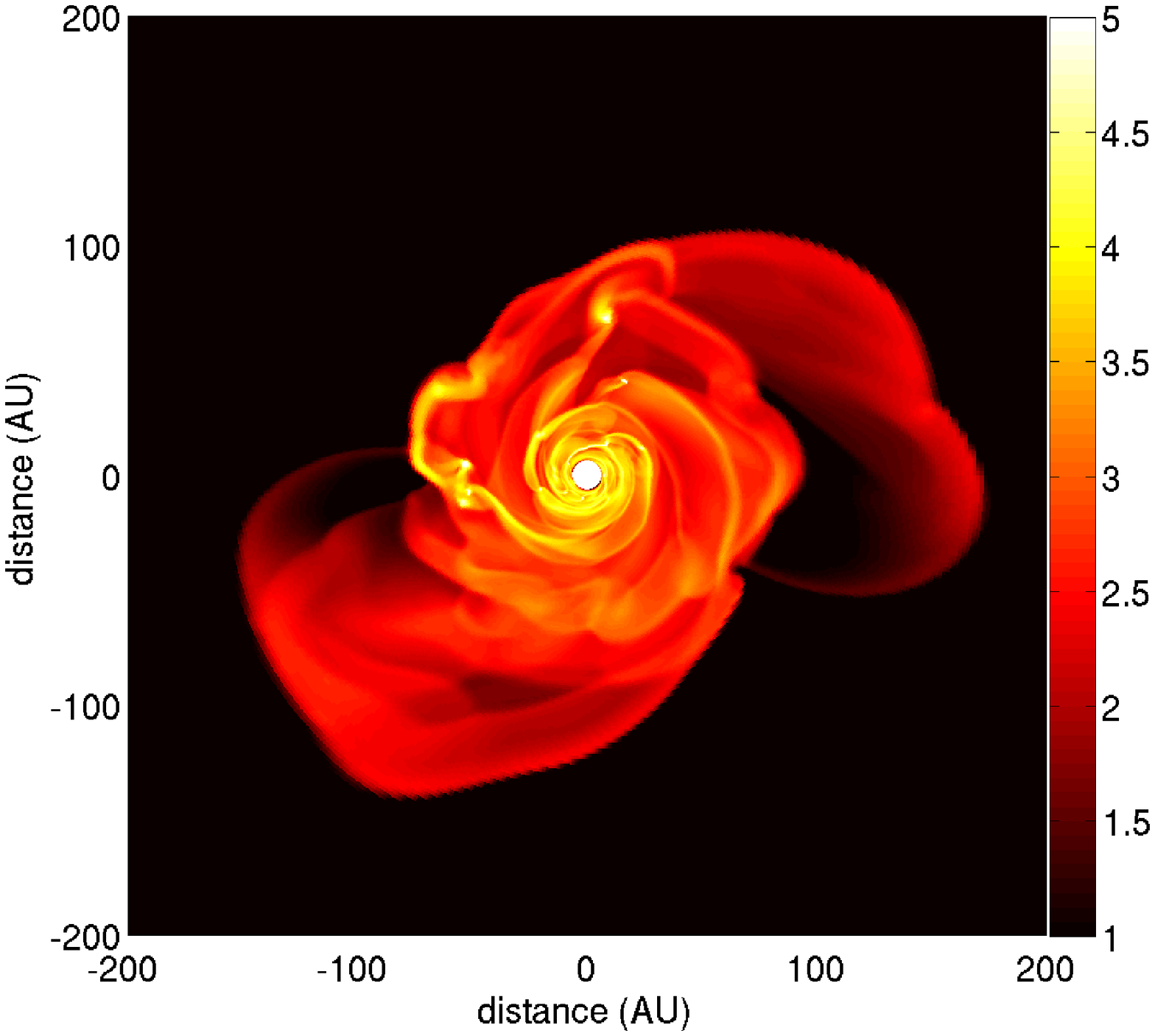}
  \includegraphics[width=0.333\textwidth]{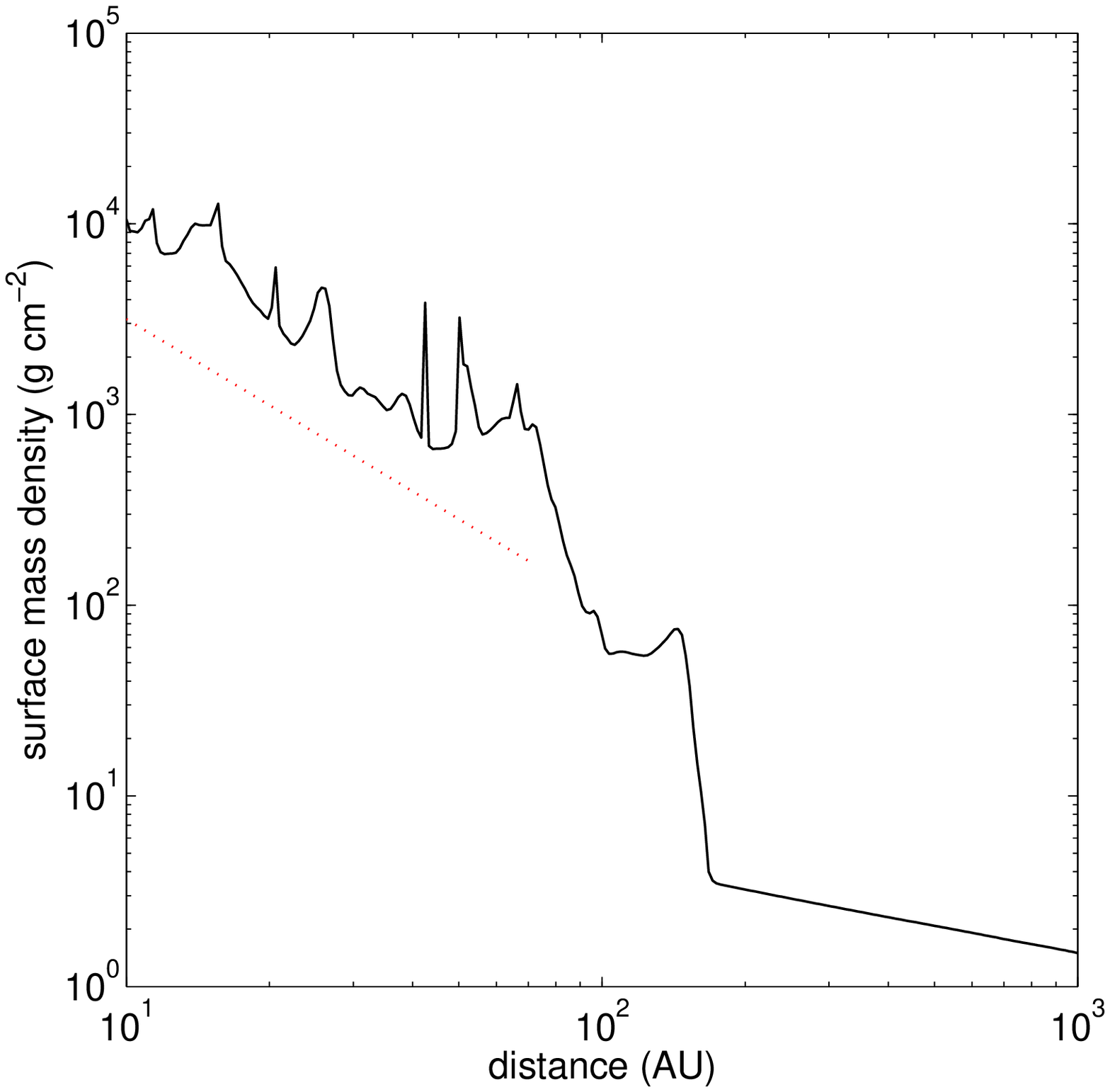}
  \includegraphics[width=0.333\textwidth]{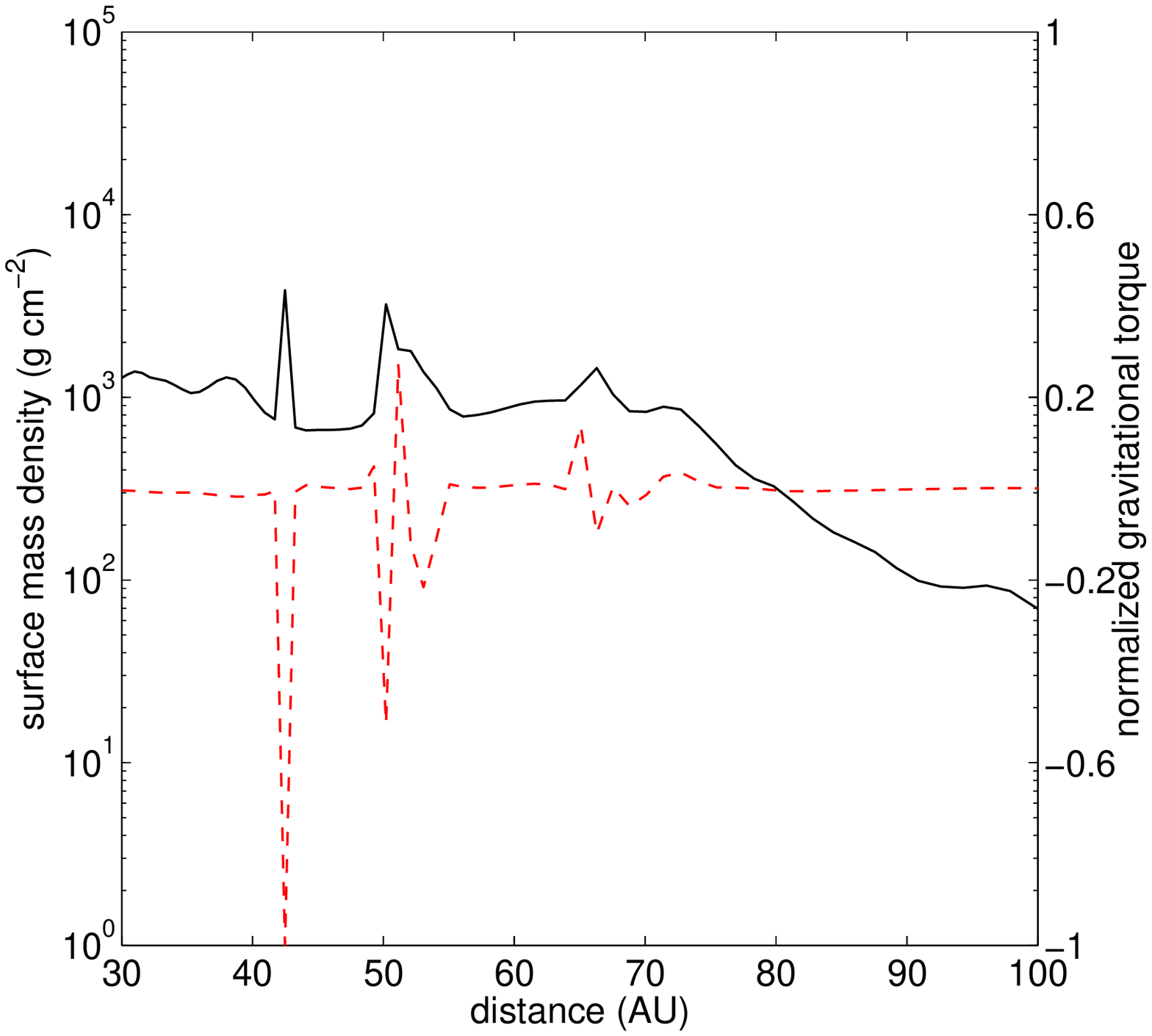}
  \caption{\textbf{Left:} Surface mass density map ${\sim}11~\mbox{kyr}$ after the formation of the primordial protostar. The gradient indicates the logarithm of the surface mass density in $\mbox{g}~\mbox{cm}^{-2}$. \textbf{Center:} Azimuthally averaged radial surface mass density profile. The dotted line shows a $r^{-1.5}$ profile for comparison. \textbf{Right:} Azimuthally averaged radial profiles of the surface mass density (solid) and normalized gravitational torque (dashed).}
  \label{fig:surfacemassdensity}
\end{figure}

We follow the evolution of isolated, gravitationally unstable, prestellar primordial cores as they collapse into the protostar and disk formation stage. The dynamics of the entire system are followed self-consistently on a single, globally defined computational domain. Here we focus on the innermost region where the disk forms around the central star. In Figure~\ref{fig:surfacemassdensity} we present a map of the surface mass density from a reference model, together with profiles of the azimuthally averaged surface mass density and gravitational torque. Within a time $t~{\leq}~5~\mbox{kyr}$ the disk begins to experience the first episodes of fragmentation. Fragments are typically formed at radial distances of ${\sim}100~\mbox{AU}$ from the central protostar, and their subsequent orbital dynamics depend on the sign of the gravitational torque acting on them. In our simulations all fragments are seen to either migrate inward, or are tidally dispersed while still in the outer disk. These outcomes result from a combination of the gravitational torques from the passage of spiral arms and interactions with other fragments. The fragmentation process is driven by the nearly continuous supply of new material infalling from the envelope.

\begin{figure}[t!]
  \includegraphics[width=0.333\textwidth]{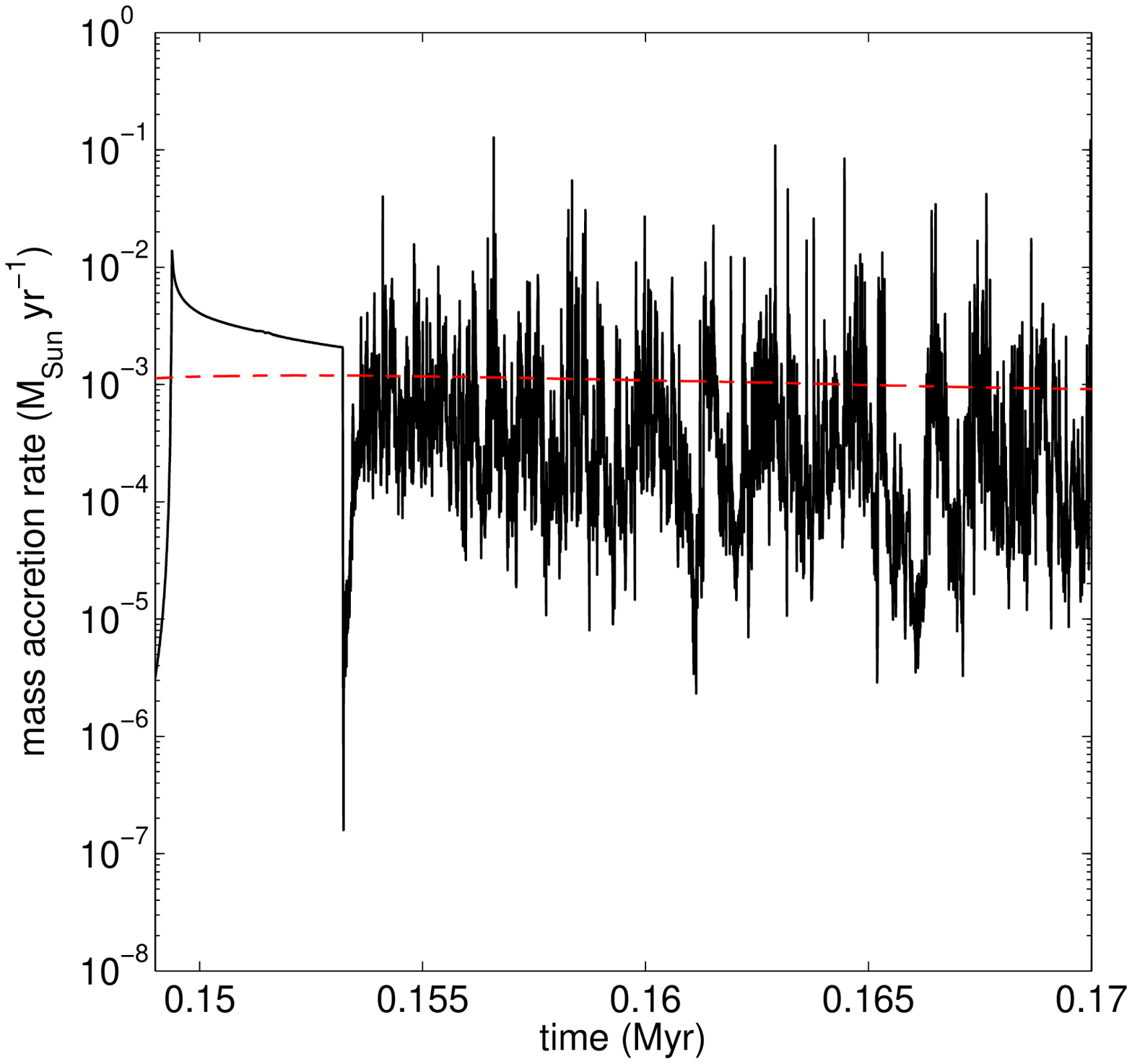}
  \includegraphics[width=0.333\textwidth]{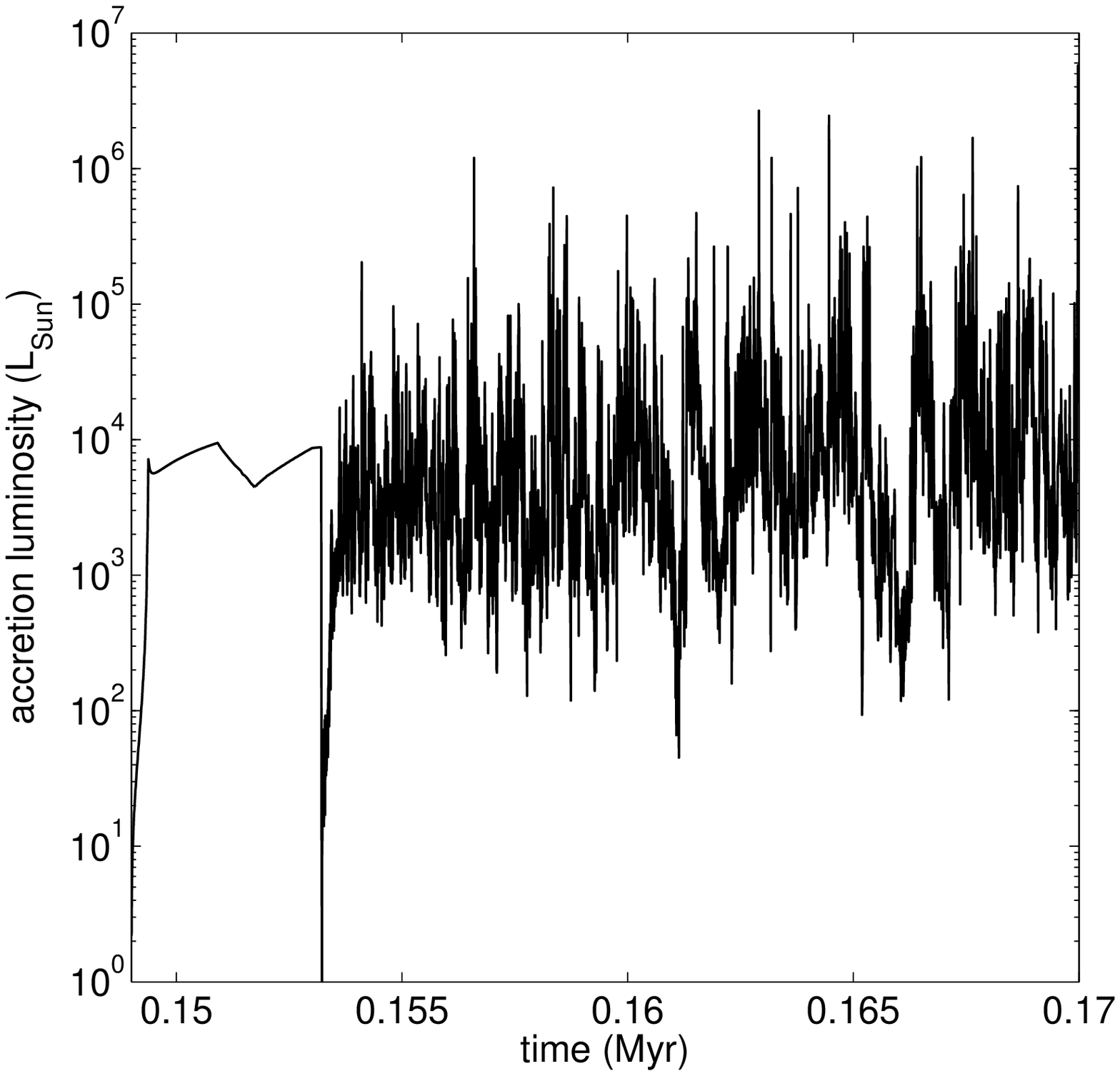}
  \caption{\textbf{Left:} Mass accretion rates, from the disk onto the central protostar (solid), and from the outer envelope onto the disk (dashed). \textbf{Right:} Accretion luminosity associated with the infall from the disk onto the central protostar.}
  \label{fig:accluminosity}
\end{figure}

Fragments surviving the inward migration and transition through the inner domain wall are likely to be tidally destroyed as they plunge into the central star, releasing their gravitational energy in the form of strong luminosity outbursts (lasting approximately 100 years). Estimating the protostellar radius with simple power-law approximations that characterize the distinct phases of the star's internal dynamics \citep{omukai2003,smith2011}, we calculate the luminosity associated with each burst of accretion (Figure~\ref{fig:accluminosity}). The luminosity generated through this burst mode of accretion is as high as a few times $10^6~\mbox{L}_{\odot}$. That the predominant fraction of material accreted onto the protostar occurs via the burst mode suggests previous estimates of the ionizing UV-flux associated with the earliest phases of the protostar's evolution \citep{mckee2008} may be underestimated by as much as a factor of $10$. Energy deposition into the environment of this order has the potential to significantly affect the nature of mass accretion onto primordial stars in the early universe.


\begin{theacknowledgments}
EIV aknowledges support from the RFBR grant 10-02-00278.
\end{theacknowledgments}

\bibliography{myrefs}


\end{document}